\documentclass[a4]{ecai2014}
\usepackage{times}
\usepackage{graphicx}
\usepackage{latexsym}

\usepackage{times}
\usepackage{helvet}
\usepackage{courier}
\usepackage{xspace}
\usepackage{amssymb}
\usepackage{amsmath}
\usepackage{epic}
\usepackage{graphicx}
\usepackage{subfig}
\usepackage[ruled,linesnumbered,boxed]{algorithm2e}
\usepackage{multirow}
\usepackage{hyperref}

\newtheorem{mytheorem}{Proposition}

{\bf}{\it}

\newcommand{\myproof}{\noindent {\bf Proof:\ \ }}
\newcommand{\myqed}{\mbox{$\blacksquare$}}

\newcommand{\myOmit}[1]{}

\newcommand{\nina}[1]{{#1}}

\begin{document}


\title{The Computational Impact of \\Partial Votes on Strategic Voting}
\sloppy



%
%
%
%

%

\author{
Nina Narodytska \institute{University of Toronto,
Canada, and  UNSW, Sydney, Australia, email: ninan@cs.toronto.edu} \and Toby Walsh \institute{NICTA and UNSW, Sydney, Australia, email: toby.walsh@nicta.com.au. NICTA is funded by
the Australian Government as represented by
the Department of Broadband, Communications and the Digital Economy and
the Australian Research Council. The author is also supported
by AOARD Grant FA2386-12-1-4056.} }

\maketitle
\bibliographystyle{ecai2014}

\begin{abstract}
In many real world elections, agents
are not required to rank all candidates.
We study three of the most
common methods used to modify voting rules
to deal with such {\em partial} votes.
These methods modify scoring rules (like the Borda count),
elimination style rules (like single transferable vote)
and rules based on the tournament graph (like Copeland)
respectively. We argue that with an elimination
style voting rule like single transferable vote,
partial voting does not change the situations where strategic
voting is possible. However,
with scoring rules 
and rules based on the tournament graph, partial voting
can increase
the situations where strategic voting is possible.
As a consequence, the computational complexity of
computing a strategic vote can change.
For example, with Borda count, the complexity of
computing a strategic vote
can decrease or stay the same depending
on how we score partial votes.
\end{abstract}


%
%
%

\section{INTRODUCTION}

Voting is a simple but general mechanism to
aggregate the preferences of multiple agents.
Much work in social choice supposes
voters declare a complete ordering over
all candidates. In practice, however, voting
systems often permit voters to declare an ordering
over a subset of the candidates.
For example, in single transferable vote elections for the
Maltese parliament,
for the Legislative Assembly
of the Australian Capital Territory,
and for the President of Ireland,
voters rank as many or as few candidates as they wish.
When all candidates in a partial vote have
been eliminated, this vote is ignored.
As a second example,
in elections for the Free Software Foundation Europe,
voters can again rank as many or as few candidates
as they wish. Unranked candidates are considered as
equal last when constructing the tournament graph used to compute
the Schulze winner. As a third example,
the Irish Green Party uses the modified Borda
count to elect its chair.
Voters can again rank as many or as few candidates
as they wish. If a voter ranks just $k$ candidates,
then their $i$th choice is given $k-i+1$ points.
The candidate with the most total points wins.

Partial voting can have a significant effect
on elections \cite{escw2012}.
For example, one reason given for the
French Academy to drop the Borda count
was ``voters had found how to manipulate
the Borda rule \ldots by truncating their lists''
(page 40 of \cite{muclassics}).
As a second example,
in elections for the 
Tasmanian
Parliament, voters are forced to rank a minimum number
of candidates to prevent certain
types of strategic voting
(for example, when three candidates are running, voters
must rank at least two candidates, whilst when four or more
candidates are running, at least three candidates must be ranked).

In this paper, we show that
partial voting has a significant
impact on computational issues surrounding
strategic voting.
Partial voting has a
similar but not completely
identical impact on related problems
like computing possible and necessary
winners, and campaign management with
truncated ballots \cite{bflraamas2012}.
{For example,
manipulating with partial votes is different to
the possible winners problem with top truncated ballots
\cite{bflraamas2012}
since manipulating votes must be complete
in the latter problem. On the other hand,
manipulating with partial
votes is equivalent to the extension bribery problem with
zero-costs. }
One important
lesson from this research is that
it would be worthwhile to re-visit
much previous work in computational social choice 
which assumes complete votes \cite{fpaimag10,fhhcacm10}.

\section{BACKGROUND}

A {\em complete vote}
is a linear order over the $m$ candidates.
We consider {\em partial votes} that are a linear
order over a strict subset of the $m$ candidates
(sometimes called ``top truncated'' votes).
An interesting extension of this work would
be to other forms of partial vote (e.g.
when voters only order a subset of
the candidate pairs).
A {\em voting rule} is a function that
maps a tuple of votes to the unique winning alternative.
We consider several common voting rules defined
on complete votes:
\begin{description}
\item[Scoring rules:] $(s_1,\ldots,s_m)$ is a vector
of scores, the $i$th candidate in a total order
scores $s_i$, and the winner is the candidate with
highest total score. The {\em plurality} rule has
the vector $(1,0,\ldots,0)$,
whilst
the {\em Borda count} has the vector $(m-1,m-2,\ldots,0)$.

\item[Single transferable vote (STV):]
This proceeds in rounds.
Unless one candidate has a majority of
first place votes, we eliminate the candidate
with the least number of first place votes.
Ballots with the eliminated candidate in first
place are re-assigned to the second place candidate.
We then repeat until a candidate has a majority.

\item[Copeland:] The candidate
with the highest Copeland score wins. The
Copeland score of candidate $i$
is $\sum_{i \neq j} (N(i,j)>\frac{n}{2})-(N(i,j)<\frac{n}{2})$
where $N(i,j)$ is
the number of voters preferring $i$ to $j$
and $n$ is the number of voters.
The Copeland winner is the candidate that wins the
most pairwise elections. Formally this is
Copeland$\mbox{}^{0.5}$ but for brevity,
we simply write Copeland.
\end{description}
We discuss in the next section how these rules
can be modified to work with partial votes.
All these rules can be easily modified to work with {\em weighted}
votes.
A vote of integer weight $w$ can be viewed as $w$ agents
who vote identically.
To ensure the winner is unique, we will
sometimes need to break ties.
A typical assumption made in the literature (and in
this paper) is
that ties are broken in favour of the manipulator.
In real world elections, ties are often broken
at random (e.g. by tossing a coin or choosing
a random vote). In this case, our results
can be seen as deciding
if we can give our preferred candidate a non-zero
chance of winning.

We will consider one agent or a coalition of agents
trying to manipulate the result of the election.
Manipulation is where these agents
vote differently to their true
preferences in order to change the outcome
whilst the other voters vote truthfully.
As in earlier work (e.g. \cite{stvhard,csljacm07}),
we consider two cases where
computational complexity may provide a shield
against manipulation: unweighted votes, a small
number of manipulators and an unbounded number
of candidates; or weighted votes, a small
number of candidates and a coalition
of manipulators of unbounded size.
%
We assume that the manipulators have {\em complete}
knowledge of the other votes. Even though this
can be unrealistic in practice, there are several reasons
why this case is interesting. First,
any computational hardness results
for complete information directly imply hardness
when there is uncertainty in the votes.
Second, results about the hardness of manipulation
by a coalition with weighted votes and complete
information imply hardness of manipulation
by an individual agent with unweighted votes and incomplete
information \cite{csljacm07}.
Third, by assuming complete information, we
factor out any complexity coming from the uncertainty
model and focus instead on computing just the manipulation.

\section{PARTIAL VOTES}

In practice, voters appear to take advantage of partial voting.
As we already noted, it was
observed that members of
French Academy cast truncated votes in an attempt
to manipulate
the Borda count.
As a second example, in the 1992 General Election for
Dublin North, 
12 candidates ran, but the
43941 voters ranked only a median of 4 candidates,
and a mean of 4.98 candidates, with
a standard deviation of 2.88 candidates.
In fact, only 8.3\% of voters cast a complete
vote.  Similarly, in the 1992 General Election for
Dublin West, 
9 candidates ran,
but the
29988 voters again ranked only a median of 4 candidates,
and a mean of 4.42 candidates, with
a standard deviation of 2.33 candidates.
In this case, 12.7\% of the voters cast a complete vote.

We consider a partial vote that ranks
just $k$ out of the $m$ candidates. 
There are a number of different ways that voting
rules can be modified to deal with partial
votes. We consider three voting
rules (Borda count, STV and Copeland) which illustrate
the most common ways to treat partial votes.
These rules allow us to cover the spectrum of
possible impacts that partial voting has on
manipulation. With scoring rules like the Borda count, we can
adjust the scoring vector to deal with
a partial vote (e.g. by shifting
it down as in the modified Borda count).
With elimination style rules like STV,
we can simply ignore votes once all their
candidates are eliminated.
Finally, with rules based on the tournament graph like Copeland,
we can simply treat unranked candidates in a partial
vote as tied in last place.
We will look at each method for dealing
with partial votes in turn.

\section{SCORING RULES}
\label{s:scoring}
The first method we study to deal with a partial
vote is to shift the scoring vector and
score unranked candidates appropriately.
Three possible schemes can be found
in the literature for dealing with
when voters rank just $k$ out of the $m$
candidates ($k<m$):
\begin{description}
\item[Round up:] A candidate ranked in $i$th place ($i\leq k$)
gets a score of $s_i$, unranked candidates get a score of 0.
For example, a partial vote that only ranks a single
candidate gives that candidate a score of $s_1$, and
0 to every other candidate. We denote this 
Borda$\mbox{}_{\uparrow}$.
\item[Round down:]
A candidate ranked in $i$th place ($i\leq k$)
gets a score of $s_{m-(k - i)-1}$, whilst
unranked candidates get a score of $s_m$.
The modified Borda count is an instance
of such rounding down.
For example, with the modified Borda count,
a partial vote that only ranks a single
candidate gives that candidate a score of $s_{m - (1-1)-1} = s_{m-1}=1$, and
0 to every other candidate.
As a second example, a partial vote that ranks
two candidates, gives the first ranked candidate
a score of $s_{m-(2-1)-1}=s_{m-2}=2$, a score of $s_{m-(2-2)-1}=s_{m-1}=1$ to the
second ranked candidate and 0 to every one else.
\nina{If $k = m$ we use Borda count to compute scores.}
\item[Average score:]
A candidate ranked in $i$th place ($i\leq k$)
gets a score of $s_i$, and unranked candidates get
$\frac{\sum_{m\geq j>k} s_j}{(m-k)}$, the average remaining
score. For example, a partial vote that only ranks one out of four
possible candidates gives that candidate a score of $s_1$, and
$\frac{s_2+s_3+s_4}{3}$,
the average of the remaining scores to the other three candidates.
We denote this 
Borda$\mbox{}_{av}$.
\end{description}
We will show that which of these three schemes we choose to
deal with partial votes can have a strong impact
on the computational complexity of computing
a manipulation.

\subsection{Borda and unweighted votes}

Partial voting increases the
situations where an election using the Borda count
can be manipulated.
For example, suppose we have three candidates ($a$, $b$ and $p$)
and a manipulator who wants $p$ to win.
One vote has been cast for each of $a>b>p$ and
$b>a>p$. With complete votes,
a manipulator cannot make $p$ win. The manipulator must
cast a vote that gives at least one point to
$a$ or $b$ thereby defeating $p$. However, with
Borda$\mbox{}_{\uparrow}$,
the manipulator can cast a vote for just
$p$ who wins by 
tie-breaking.
Partial voting can also change the computational
complexity of computing a manipulation.
With complete votes, computing if two
voters can manipulate the Borda count
is NP-hard \cite{dknwaaai11,borda2}.
On the other hand, with partial voting and rounding
up, computing such a manipulation takes
polynomial time. 

\begin{mytheorem}\label{t:bordaup}
Computing if a coalition of manipulators can manipulate
Borda$\mbox{}_{\uparrow}$ with unweighted
and partial votes takes polynomial time.
\end{mytheorem}
\myproof
The manipulators simply vote for the
candidate who they wish to win and no one else.
This is the best possible vote.
\myqed

If we treat partial votes by
rounding down or averaging the
remaining scores, computing a
manipulation remains 
intractable.

\begin{mytheorem}\label{t:mbc_two}
Computing if two voters can manipulate
the modified Borda count
or Borda$\mbox{}_{av}$ with unweighted
and partial votes is NP-hard.
\end{mytheorem}
\myproof
We use the same reduction as in
\cite{dknwaaai11}. To ensure
that the preferred (first) candidate with an initial score
of $C$ wins and that the $n+2$th (of $n+3$) candidate
with an initial score of $2(n+2)+C$ does not,
the two manipulators must cast
\nina{a complete
vote for all $n+3$ candidates with
their preferred candidate in the first place,
and the $n+2$th candidate in the last place for Borda$\mbox{}_{av}$.
If we use the modified Borda count, manipulators can also cast partial votes of length $n+2$ with
their preferred candidate in the first place and the $n+2$th candidate not ranked. This also
achieves the manipulators' goal of reducing the gap between the preferred candidate and the $n+2$th candidate to 0.}
Hence,
partial voting does not increase the
ability of the manipulators to manipulate
the problem instances used in the reduction.
\myqed

\subsection{Borda and weighted votes}

We now turn to weighted votes.
With complete votes, computing a
coalition manipulation of the Borda count with just 3
candidates is NP-hard \cite{csljacm07}.
With partial votes and rounding up, computing
such a manipulation now takes polynomial time.

\begin{mytheorem}
Computing a coalition manipulation of
Borda$\mbox{}_{\uparrow}$ with weighted
and partial votes takes polynomial time.
\end{mytheorem}
\myproof
The coalition simply votes for their most
preferred candidate and no one else.
\myqed

On the other hand, if we treat partial votes by
rounding down or averaging the
remaining scores, computing a coalition
manipulation remains computationally intractable.

\begin{mytheorem}
Computing a coalition manipulation of
the modified Borda count with weighted
and partial votes and 3 candidates is NP-hard.
\end{mytheorem}
\myproof
Reduction from the number partitioning problem.
We have a bag of integers, $k_i$ summing to $2K$
and wish to decide if we can
divide them into two partitions,
each with sum $K$.
We consider an election over three candidates,
$a$, $b$ and $p$ in which the manipulating
coalition wish $p$ to win.
\nina{We have partial votes of weight $3K$
for $a$ and for $b$.
Hence, the score of $a$ is $3K$
and the score of $b$ is $3K$.}

The voters in the coalition each have
weight $k_i$. We identify voters
in the coalition by the corresponding integer
$k_i$. Suppose a partition exists.
\nina{Let those manipulators in
one partition vote $p>a>b$
and the others vote $p>b>a$.
Now, $a$, $b$ and $p$ all have scores of $4K$
so $p$ wins by tie breaking.
Conversely, suppose $p$ wins.
We can suppose no manipulator
votes for just $a$ or just $b$
(as this is counter-productive).
Suppose the manipulators
have votes of weight $x$ for $p>a>b$,
$y$ for $p>b>a$ and $z$ for just $p$.
Aside: a vote for $p>a$ is the same as one
for $p>a>b$ as only $p$ gets two points
and $a$ gets one point in these votes,
similarly a vote for $p>b$ is the same as one for $p>b>a$.}
Now $x+y+z=2K$.
Since $p$ wins, $p$ beats $a$.
That is, $2(x+y)+z \geq 3K+x$.
This simplifies to $(x+y+z)+y \geq 3K$.
Substituting $x+y+z=2K$ gives
$y \geq K$.
Similarly, $p$ beats $b$. That
is $2(x+y)+z \geq 3K+y$.
Again this gives $x\geq K$.
But $z = 2K-x-y$. Hence, $z \leq 0$.
Thus, $x=y=K$ and $z=0$. That is, we
have a perfect partition.
Note that
the proof in \cite{csaaai2002}
showing that coalition manipulation of
the Borda count with weighted and complete votes
is NP-hard does not work
for the modified Borda count. In
this reduction, the final scores
(of $24K$ and $24K-3$) are not
close enough to preclude a manipulation
using both complete and partial votes
even when there is no perfect partition.
\myqed

For Borda$\mbox{}_{av}$,
computing a coalition manipulation
with partial votes is also NP-hard.
We have a relatively simple proof
for 4 candidates based on a reduction
from number partitioning similar to
that for the modified Borda count. For 3 candidates,
our proof is much more complex and
requires reduction from a very specialized
subset sum problem which
we prove is itself NP-hard.

\begin{mytheorem}
Computing a coalition manipulation of
Borda$\mbox{}_{av}$
with weighted
and partial votes and 3 candidates is NP-hard.
\end{mytheorem}
\myproof
The proof uses a reduction from a
specialized subset sum problem with
repetition.
Given a bag of positive integers $S$,
such that $S$ can be partitioned
into pairs of identical numbers,
and a target sum $t$, we consider
deciding if there is a subset $S'$ of $S$ whose sum is $t$.
To show that this subset sum problem is NP-hard,
we modify the
reduction of 3SAT to subset sum in~\cite{3sat2subsetsum}.
Consider a CNF formula with $n$ variables and $m$ clauses.
For each literal $x_i$ and $\bar{x}_i$ we introduce two equal numbers,
$y_i,y'_i$, $y_i = y'_i$, and $z_i,z'_i$, $z_i = z'_i$, $i=1,\ldots,n$, respectively.
For each clause $C_j$ we introduce two equal numbers,
$g_j$ and $g'_j$, $j=1,\ldots,m$.
By construction, it follows that numbers in $S$
can be partitioned into pairs of identical numbers.
Each number $y_i, y'_i, z_i, z'_i, g_j$ and $g'_j$
is a decimal number with $n+m$ digits. We call
the first $n$ digits variable-digits and
the last $m$ digits clause-digits.
Consider the $y_i$ number, $i=1,\ldots,n$.
The $i$th digit in $y_i$ is one.
If $C_j$ contains $x_i$ then the $(n+j)$th digit is 1.
The remaining digits are zeros.
The $y'_i$ number is identical to $y_i$, $i=1,\ldots,n$.
Similarly, we define numbers $z_i(z_i')$, $i=1,\ldots,n$.
The $i$th digit in $z_i$ is one. If $C_j$ contains $\bar{x}_i$ then
the $(n+j)$th digit is 1. The remaining digits are zeros.
Consider numbers $g_j(g_j')$, $i=1,\ldots,m$.
The $(n+j)$th digit is 1. The remaining digits are zeros.
Finally, we introduce the target number $t$.
The first $n$ digits equal one and the last $m$ digits
equal $3$.

\textbf{Assignment encoding.} As first
$n$ variable-digits of $t$ are ones, only one of the
numbers $y_i$, $y'_i$, $z_i$, $z'_i$ can be selected
to $S'$. Hence, selection of $y_i$ or $y'_i$ to $S'$
encodes that $x_i = 1$, and a selection of $z_i$ or $z'_i$ to $S'$
encodes that $x_i=0$.

\textbf{Checking an assignment.}
Last $m$ clause-digits of $t$ equal $3$. Consider
a clause $C_j = (x_i, \bar{x}_s, x_k)$.
If none of the variables $y_i,y'_i,z_s,z'_s, y_k$ and $y'_k$
is selected to the set $S'$ then the maximum value
in the $(n+j)$th digit is two. Hence, one of these variables
must be selected.
The reverse direction is trivial.
Hence, this subset sum problem with repetition is NP-hard.

We use this problem to show NP-hardness of
coalition manipulation of
Borda$\mbox{}_{av}$
with 
3 candidates.
Given a set of positive integers $S = \{s_1,s'_1\ldots, s'_n,s_n\}$, such that all elements of $S$ can be partitioned
into pairs of identical numbers, $\{s_i,s'_i\}$, $i=1,\ldots,n$, and a target sum $t_1$, we consider if
there is a subset of $S$, $S'$, whose sum is $t_1$.
We assume that $t = \sum_{s_i \in S} (s_i +s'_i)$. We denote $t_2 = t - t_1$.
We have an election over three candidates ($a$, $b$ and
$p$) in which the manipulating coalition wish $p$ to win.
We have one complete vote of weight $ t_1$
for $a>b>p$ and  one complete vote of weight $t_2$
$b>a>p$. The total scores from non-manipulators are
$score(a) = 2t_1 + t_2 = t_1+t$,
$score(b) = 2t_2 + t_1 = t_2+t$
and
$score(p) = 0$.
The voters in the coalition each have
weight $(s_i + s'_i)$. Suppose a subset sum $S'$ exists.
Consider three cases.
If $s_i$ and $s'_i$ are in $S'$ then the $i$th
manipulator votes $p >b>a$.
If $s_i$ and $s'_i$ are not in $S'$ then the $i$th
manipulator votes $p>a>b$.
If $s_i$ is in and $s'_i$ is not in $S'$ then the $i$th
manipulator votes $p$. Hence, $a$ and $b$ get $s_i=s'_i$
points each. The case when $s_i$ is not in and $s'_i$ is in $S'$
is similar.
As $S'$ exists, the score of $b$ from manipulators
is exactly the sum of numbers in $S'$ which is
equal to $t_1$. The preferred candidate $p$ gets
$2t$ points which is the sum of all elements in $S$ multiplied by $2$.
Finally, $a$ gets $t_2$ points
which is the sum of all elements in $S \setminus S'$.
Hence, the total scores are
$score(a) = t_1+t +t_2 = 2t$,
$score(b) = t_2+t+t_1 =2t$
and
$score(p) = 2t$.
The  preferred candidate $p$ wins by the tie-breaking rule.

Conversely, suppose $p$ wins.
We show that $p$ wins iff $p$ is ranked first
in all manipulators votes.
Suppose $p$'s score is $2t-\epsilon$, $\epsilon > 0$, so that it is not ranked
first in all manipulator votes. Hence, $a$ and $b$
have to share $t+\epsilon$ points between them as we have
$3t$ points to distribute. Let $a$ get  $q_1$
and $b$ get $q_2$ points out of $t+\epsilon$ points,
$q_1 + q_2 = t+\epsilon$.
For $p$ to be a co-winner the following must hold:
$t + t_1 + q_1 \leq 2t-\epsilon$ and
$t + t_2 + q_2 \leq 2t-\epsilon$.
If we sum up these two inequalities we get
$2t + (t_1+t_2) + (q_1+q_2) = 4t -\epsilon \leq 4t-2\epsilon$.
This leads to a contradiction.
Therefore, $p$ is ranked first in all manipulators votes.
In this case, there are exactly $t$ points that
the manipulators have to distribute between $a$ and $b$.
Let $a$ get $q_1$
and $b$ get $q_2$ points out of $t$ points,
$q_1 + q_2 = t$. We also know that
$t + t_1 + q_1 \leq 2t$ and
$t + t_2 + q_2 \leq 2t.$
Hence,  $q_1 \leq t-t_1 = t_2$ and  $q_2 \leq t-t_2 = t_1$.
As $q_1 + q_2 = t$, $q_1 = t_2$ and $q_2=t_1$.
In a successful manipulation there are three types
of votes: $ p > a > b$,   $ p > b > a$ and $p$.
If the $i$th manipulator votes $p > b > a$ then $b$
gets $s_i+s'_i$ points and we say that $s_i$ and $s'_i$ belong to $S'$.
If the $i$th manipulator  votes $p$ then $b$ gets $s_i$ points and we say that
$s_i$ belongs to $S'$. If the $i$th manipulator
votes $p>a>b$ then $b$ gets $0$ points.
As $b_1$ gets exactly $t_1$ then the sum of the numbers
in $S'$ is exactly $t_1$.
\myqed

For other scoring rules besides the Borda count,
it appears likely that similar results
can be given for the impact of partial
voting on weighted and unweighted manipulation.

\section{SINGLE TRANSFERABLE VOTE}

We now consider the second type of method for
dealing with partial votes.
For elimination style rules like STV, a method
analogous to rounding up for
scoring rules is used in many real
world settings. We simply ignore
a partial vote once all the candidates
in the vote have been eliminated.
Unlike with the Borda count,
partial voting in STV elections does not
permit more manipulations to take place.

\begin{mytheorem}
Under STV, if a coalition of agents
can cast partial votes to ensure a given candidate wins
then they can also cast complete votes for the same
outcome.
\end{mytheorem}
\myproof
Suppose the agents can cast partial votes to
ensure a given candidate $p$ wins.
We can complete each of their votes without
changing the outcome of the election.
We simply add $p$ to the end of
the partial vote (if it does not
already include $p$). Then we add
the remaining candidates in
any order. Such a completion does not
change the result.
If the partial vote included
$p$, then the completion will never
be considered.
If the partial vote didn't include $p$,
then we
have merely added another vote for $p$
from the point that all the candidates
in the partial vote have been eliminated.
This only helps $p$ to win.
\myqed

Since partial voting does not change
which elections can be manipulated, it follows
immediately that
the computational complexity
of computing a manipulation of STV
remains unchanged when we permit
partial voting. In particular, with weighted
votes, computing a
coalition manipulation of STV with 3
candidates and complete votes is NP-hard \cite{csljacm07}.
The problem remains computationally intractable
when the manipulating coalition can cast partial votes.
Similarly, with unweighted votes, it is NP-hard
for a single agent to compute a
strategic manipulating vote of
STV \cite{stvhard}.
The problem again remains computationally intractable
with partial voting.
It would be interesting to identify
other voting rules where partial
voting has no impact on manipulation.
Not all
elimination style rules are unchanged by
partial voting. For instance, it is easy to
see that Borda style elimination rules like
Nanson and Baldwin are impacted by
partial voting.
\myOmit{On the other hand,
an elimination style voting rule like Coombs' (in which
we eliminate the candidate with the most vetoes)
are not impacted by partial voting. We suppose here that
voters are able to give a partial vote
that lists only their $k$ least preferred candidates. }

\section{TOURNAMENT GRAPH RULES}

We now consider the third method for
dealing with partial votes.
For voting rules based on the tournament graph like Copeland,
a method for dealing with partial
votes analogous to rounding up for
scoring rules is used in several real
world settings. More particularly,
we consider unranked candidates
to be tied in last place.
Such partial voting increases
the situations where manipulation
is possible.
Suppose we have 4 candidates: $a$, $b$, $c$ and $p$.
One vote each has been cast for $a>b>c>p$,
$b>c>a>p$, and $p>c>a>b$.
We have one manipulator who wants $p$ to win.
If the manipulator casts a partial vote
that just ranks $p$ in first place then every candidate
has a Copeland score of 0, and $p$
wins by tie breaking. Hence, there is
a successful manipulation with partial voting.
On the other hand, suppose the manipulator
must cast a complete vote. Now $a$, $b$ and $c$
are symmetric. They each tie with $p$ (supposing
$p$ is ranked in first place by the manipulator),
and in the fixed votes, each beats one
candidate and is beaten by one other
candidate. Without loss of generality,
we can suppose therefore that the manipulator
casts the complete vote $p>a>b>c$. In this
case, $a$ wins with a Copeland score of 1.
Hence, with complete voting, manipulation
is not possible.

\subsection{Copeland and unweighted votes}

With complete votes, a simple greedy method
will compute a strategic vote for a single
agent to manipulate the result of Copeland's
method in polynomial time when this is possible
\cite{bartholditoveytrick}. We can
adapt this method to construct a
strategic {\em partial} vote.
Our adaptation adds an additional stopping
condition which exits the procedure early with
a successful partial vote.
We suppose, as before,
that we break ties in favour of the manipulator.
It is, however, easy to relax this assumption.
The initial step of the greedy
manipulation procedure is to rank
the preferred candidate $p$
in first place. We then repeat the
following steps. If
the Copeland score of $p$ is greater
than or equal to the current Copeland scores
of all the other candidates, we stop as we have
a (possibly partial) vote with the desired outcome.
Alternatively, we still need to reduce the Copeland
scores of one or more ``dangerous'' candidates by
voting for a ``harmless'' candidate.
To do this, we determine if there
is a candidate who can be placed in the next
position in the partial vote without
giving this candidate a Copeland score
exceeding that of $p$.
We add this
candidate to the partial vote and repeat.
If there is no such candidate, then we terminate
as $p$ cannot win.

\begin{mytheorem}
For Copeland's method,
there is a greedy manipulation procedure which finds
a strategic partial vote in polynomial time
that makes a given candidate
win whenever this is possible.
\end{mytheorem}
\myproof
Suppose the procedure fails
but there exists a partial
vote $\Pi$ that makes the given candidate
$p$ win.
Consider the highest candidate $c$ 
not appearing in the partial
vote constructed by the greedy manipulation
procedure before it failed. 
If we add $c$ to this
partially constructed vote then $c$ has
a lower Copeland score than if we added the vote $\Pi$. Hence,
there was a candidate who could be harmlessly placed in
the next position in the vote. The greedy manipulation
procedure should not therefore have terminated unsuccessfully.
\myqed

\subsection{Copeland and weighted votes}

With complete votes, it is NP-hard to
compute a weighted coalition manipulation of Copeland's
method. As we argued earlier,
partial voting increases our
ability to manipulate such elections.
However, it remains computationally intractable
to compute such a manipulation.

\begin{mytheorem}
Computing a coalition manipulation of
Copeland's method with weighted and partial
votes and 4 candidates is NP-hard.
\end{mytheorem}
\myproof
Reduction from number partitioning.
We are given a bag of integers $k_i$ with sum $2K$
and wish to determine if there is a partition
into two bags each of sum $K$.
There are 4 candidates, $a$, $b$,
$c$ and $p$ where $p$ is the candidate
that the manipulating coalition prefers
to win. We suppose there are $K$ fixed identical
votes for $a>b>c>p$ and
$K$ for $a>c>b>p$. The manipulating
coalition has a voter of weight $k_i$
for each integer in the bag being
partitioned. Suppose there is a perfect partition,
and the voters corresponding to one
partition vote $p>b>c>a$
and in the other vote $p>c>b>a$.
Then all candidates have a Copeland
score of 0, and $p$ wins by the tie-breaking
rule.
On the other hand, suppose that the manipulating coalition
can vote so that $p$ wins. Without loss
of generality, we can suppose that they
all rank $p$ first. This is
the best possible outcome for $p$,
giving $p$ a Copeland score of 0. The
manipulating coalition cannot therefore cast votes that
result in $a$, $b$ or $c$ having a Copeland
score greater than 0.
Now, without the votes of
the manipulating coalition,
$a$ has a Copeland score of 3.
The manipulating coalition must all
prefer $b$ and $c$ to $a$ to
reduce this score. Hence every member of
the manipulating coalition must rank $b$ and $c$. Finally,
$b$ and $c$
are tied before the manipulating coalition
votes. If $b$ is preferred to $c$
overall then $b$ has a Copeland score
of 1. Similarly,
if $c$ is preferred to $b$
overall then $c$ has a Copeland score
of 1. Hence, $b$ and $c$ must tie. This
is only possible if the manipulating
coalition cast votes of weight $K$
for $b>c$ and of weight $K$ for
$c>a$. Thus, the manipulating
coalition must cast complete votes of weight $K$
for $p>b>c>a$ and of weight $K$ for
$p>c>b>a$.
\myqed

Note that we cannot use the reduction
used in proving the NP-hardness of
coalition manipulation of Copeland's method
with complete votes and 4 candidates
\cite{csljacm07}.
By casting a partial
vote for just $p$ and leaving all
other candidates unranked,
the manipulating
coalition can make the preferred candidate
$p$ win in this reduction even if there is not a perfect
partition.
This proof 
also requires that we break ties against the manipulating
coalition whilst our proof makes the (more common?)
assumption that we break ties in favor of the manipulating
coalition.
With just 3 candidates and tie breaking in favour
of the manipulators,
coalition manipulation of Copeland's method
with weighted and complete votes is NP-hard
(Theorem 4.1 in \cite{copeland}). Unfortunately, the reduction
used in this proof fails for partial votes.
We conjecture that the problem of computing
a manipulation of Copeland's method is NP-hard
with partial votes and 3 candidates. However, any proof
looks as involved as that required for Borda$\mbox{}_{av}$.

Our results for Borda, STV and Copeland voting rules with complete and  partial votes are summarised  in Table~\ref{table:bsc}.

\begin{table}
  \centering
  \scriptsize
  \begin{tabular}{|c|c|c|}
     \hline
      & Unweighted CM & Weighted CM \\
      \hline
            \hline
 \multicolumn{3}{|c|}{Complete votes} \\
      \hline
            \hline
     Borda  & NP-hard & NP-hard \\
     STV  & NP-hard & NP-hard \\
     Copeland  & P & NP-hard \\
      \hline
      \hline
 \multicolumn{3}{|c|}{Partial votes} \\
      \hline
            \hline
     Borda$\mbox{}_{\uparrow}$ & P & P \\
     Modified Borda & NP-hard & NP-hard \\
     Borda$\mbox{}_{av}$ & NP-hard & NP-hard \\
     STV  & NP-hard & NP-hard \\
          Copeland  & P & NP-hard \\
     \hline
   \end{tabular}
  \caption{Summary of results.}\label{table:bsc}
\end{table}

\section{INTRODUCING INTRACTABILITY}

We have seen that
partial voting has a range of effects on the
computational complexity of computing a manipulation.
\begin{enumerate}
\item
Partial voting does not change when
strategic voting is possible,
and thus there is no change also in
the computational cost of
computing a strategic vote (e.g. STV).
\item Partial voting permits more
strategic voting but there is
no change in the worst case complexity of
computing a strategic vote
(e.g. the modified Borda count).
\item Partial voting permits more
strategic voting and
the worst case complexity of
computing a strategic vote decreases
(e.g. Borda$\mbox{}_{\uparrow}$)
\end{enumerate}
We now demonstrate the fourth and final
possibility: partial voting permits more
strategic voting and
the worst case complexity of
computing a strategic vote increases.
This occurs when 
a strategic
but complete vote takes polynomial time to
compute whilst a strategic but partial
vote is NP-hard to compute.
In fact, our
proof that demonstrates there exists a {\em subclass}
of elections where computing a manipulation with
complete votes takes polynomial time (because it is never
possible), but with partial votes and two
manipulators it is NP-hard.

\begin{mytheorem}
There exists a variant of
Borda voting, and a class of
elections where it takes polynomial time for two agents
to compute their strategic vote when they
must cast complete votes but it is NP-hard
with partial votes.
\end{mytheorem}
\myproof
We consider the scoring rule in which
a candidate ranked in $i$th position
gets a score of $m-i+2$
where $m$ is the total number
of candidates. Hence the last ranked
candidate in a complete vote gets a score of $2$.
With partial votes, we suppose scores are rounded down.
That is, if only $k$ candidates
are ranked, then the $i$th ranked
candidates gets a score of $k-i+2$,
and unranked candidates get a score of
$0$. We adapt the reduction used in \cite{dknwaaai11}.
We add one ``dangerous'' candidates
(to give $n+4$ candidates in total).
This candidate gets
a score from the fixed votes of $C+2(n+4)$.
All other candidates get the same score as
in the reduction in \cite{dknwaaai11}. Now,
if either of the manipulating agents casts
a complete vote, the dangerous candidate
increases their score so is sure to win.
In fact, the only way for the dangerous candidate
not to win is for both manipulating agents
to cast a partial vote ranking all but
the dangerous candidate. With such a
vote, the dangerous candidate will draw with
the other leading scorers. The proof then
follows the same argument as in \cite{dknwaaai11}.
\myqed

\section{PARTIAL VOTING IN PRACTICE}
\begin{table*}[ht]
\centering
\scriptsize
\begin{tabular}{|l@{}c@{}c|c@{}c|@{}c@{}c|@{}c@{}c|@{}c@{}c||c@{}c|@{}c@{}c|@{}c@{}c|@{}c@{}c|}
\hline
problem name,\#id& $m$& $t$& \multicolumn{8}{|c||} { Borda$\mbox{}_{\uparrow}$} & \multicolumn{8}{|c|} { Modified Borda count}\\
\hline
\hline
& &&\multicolumn{6}{c|} {Manipulators' votes of length:}& \multicolumn{2}{c||} {Full votes   }& \multicolumn{6}{c|} {Manipulators' votes  of length:}& \multicolumn{2}{c|} {Full votes }\\
& &&\multicolumn{2}{c} {15}& \multicolumn{2}{c} {30}& \multicolumn{2}{c|} {45}& \multicolumn{2}{c||} {  }& \multicolumn{2}{c} {15}& \multicolumn{2}{c} {30}& \multicolumn{2}{c|} {45}& \multicolumn{2}{c|} {}\\
\cline{4-19}
&&&avg t&  ~~avg p &avg t&  ~~avg p &avg t&  ~~avg p &avg t&  ~~avg p &avg t&  ~~avg p &avg t&  ~~avg p &avg t&  ~~avg p &avg t&  ~~avg p \\
\hline
F1 and Skiing,\#1 &54 &32 &  \textbf{11.8} & 34.0 &  7.2 & 16.7 &  7.4 & 11.2 &  8.1 & 9.3 &  6.5 & 18.3 &  8.3 & 18.3 &  11.1 & 16.2 & \textbf{15.7} & 17.5 \\
F1 and Skiing,\#1 &54 &64 &  \textbf{26.4} & 63.6 &  13.6 & 31.0 &  12.8 & 20.6 &  14.7 & 17.2 &  12.9 & 34.7 &  16.4 & 34.7 &  17.2 & 24.9 & \textbf{21.9} & 26.1 \\
F1 and Skiing,\#14& 62 &32 &  \textbf{17.4} & 39.0 &  10.1 & 19.1 &  9.1 & 12.8 &  11.1 & 9.3 &  8.1 & 18.6 &  10.0 & 18.6 &  14.4 & 18.6 & \textbf{17.7} & 13.4 \\
F1 and Skiing,\#14& 62 &64 &  \textbf{41.6} & 79.8 &  20.7 & 38.8 &  18.2 & 25.8 &  28.5 & 18.6 &  16.4 & 37.6 &  20.1 & 37.6 &  \textbf{28.5} & 37.6 & 26.1 & 21.7 \\
F1 and Skiing,\#17& 61 &32 &  \textbf{15.3} & 36.2 &  8.8 & 17.8 &  8.5 & 11.9 &  10.2 & 8.8 &  7.8 & 18.5 &  10.0 & 18.5 &  14.2 & 18.5 & \textbf{15.1} & 11.2 \\
F1 and Skiing,\#17& 61 &64 &  \textbf{33.0} & 65.9 &  16.7 & 32.1 &  15.1 & 21.4 &  17.2 & 15.8 &  15.1 & 34.1 &  19.2 & 34.1 &  \textbf{28.6} & 33.9 & 15.9 & 14.0 \\
Sushi Data,\#2 &100 &~~32 &  4.7 & 5.1 &  4.4 & 2.7 &  4.8 & 1.9 &  \textbf{8.3} & 1.0 &  7.6 & 10.8 &  10.1 & 10.8 &  14.1 & 10.8 & \textbf{62.1} & 10.8 \\
Sushi Data,\#2 &100 &~~64 &  6.0 & 8.1 &  5.3 & 4.2 &  5.6 & 2.9 &  \textbf{11.6} & 1.7 &  12.4 & 17.0 &  16.8 & 17.0 &  23.2 & 17.0 & \textbf{91.3} & 16.9 \\
\hline
\hline
& &&\multicolumn{6}{c|} {Manipulators' votes of length:}& \multicolumn{2}{c||} {Full votes   }& \multicolumn{6}{c|} {Manipulators' votes  of length:}& \multicolumn{2}{c|} {Full votes }\\
& &&\multicolumn{2}{c} {3}& \multicolumn{2}{c} {6}& \multicolumn{2}{c|} {9}& \multicolumn{2}{c||} {  }& \multicolumn{2}{c} {3}& \multicolumn{2}{c} {6}& \multicolumn{2}{c|} {9}& \multicolumn{2}{c|} {}\\
\cline{4-19}
&&&avg t&  ~~avg p &avg t&  ~~avg p &avg t&  ~~avg p &avg t&  ~~avg p &avg t&  ~~avg p &avg t&  ~~avg p &avg t&  ~~avg p &avg t&  ~~avg p \\
\hline
Debian Project Data,\#4 &8 &32 &  6.6 & 32.8 &  4.4 & 12.7 &  &  &  \textbf{26.4} & 6.8 &  0.8 & 11.3 &  39.8 & 8.8 &  &  & \textbf{52.4} & 8.7 \\
Debian Project Data,\#4 &8 &64 &  \textbf{42.2} & 62.5 &  5.1 & 21.8 &  &  &  31.2 & 10.3 &  2.4 & 20.7 &  21.8 & 12.7 &  &  & \textbf{22.8} & 10.6 \\
Irish Election Data,\#1 &9 &32 &  1.7 & 13.6 &  0.9 & 5.7 &  &  &  \textbf{15.1} & 3.7 &  0.9 & 8.7 &  0.9 & 7.6 &  &  & \textbf{8.1} & 7.5 \\
Irish Election Data,\#1 &9 &64 &  3.5 & 24.7 &  1.1 & 10.2 &  &  &  \textbf{5.7} & 6.1 &  1.1 & 14.9 &  1.2 & 12.0 &  &  & \textbf{26.9} & 11.5 \\
Debian Project Data,\#5 &9 &32 &  4.6 & 23.6 &  1.2 & 9.8 &  &  &  \textbf{7.1} & 5.2 &  0.7 & 7.2 &  1.4 & 6.1 &  &  & \textbf{10.5} & 5.9 \\
Debian Project Data,\#5 &9 &64 &  \textbf{25.8} & 48.4 &  2.2 & 19.5 &  &  &  2.5 & 7.4 &  1.1 & 14.2 &  12.8 & 9.3 &  &  & \textbf{15.2} & 9.1 \\
Glasgow City Council,\#1 &9 &32 &  \textbf{1.1} & 8.6 &  0.7 & 3.8 &  &  &  0.9 & 2.6 &  0.8 & 7.2 &  \textbf{12.4} & 7.0 &  &  & 11.5 & 6.6 \\
Glasgow City Council,\#1 &9 &64 &  1.9 & 14.4 &  0.8 & 6.1 &  &  &  \textbf{5.7} & 4.0 &  1.0 & 13.1 &  8.3 & 11.1 &  &  & \textbf{14.4 }& 10.2 \\
ERS Data,\#1 &10& 32 &  \textbf{1.8} & 16.5 &  0.9 & 6.9 &  1.0 & 4.5 &  1.0 & 4.1 &  1.0 & 10.7 &  1.3 & 10.7 &  \textbf{7.2} & 10.5 & 7.1 & 10.5 \\
ERS Data,\#1 &10& 64 &  \textbf{3.8} & 29.3 &  1.3 & 12.0 &  1.1 & 7.7 &  1.3 & 7.0 &  1.4 & 18.2 &  2.5 & 18.1 &  \textbf{17.8} & 16.8 & \textbf{17.8} & 16.8 \\
ERS Data,\#38 &11& 32 &  \textbf{2.9} & 24.3 &  1.2 & 10.0 &  1.0 & 6.4 &  1.0 & 5.3 &  0.9 & 9.6 &  1.1 & 9.6 &  3.3 & 9.5 & \textbf{7.0} & 9.6 \\
ERS Data,\#38 &11 &64 &  \textbf{9.0} & 47.9 &  1.8 & 19.4 &  1.6 & 12.3 &  1.6 & 9.9 &  1.5 & 18.2 &  1.7 & 18.2 &  \textbf{8.8} & 18.2 & 3.7 & 18.2 \\
Irish Election Data,\#2 &12& 32 &  1.5 & 16.6 &  1.0 & 6.9 &  1.1 & 4.5 &  \textbf{18.4} & 3.4 &  1.1 & 9.6 &  1.9 & 9.5 &  6.9 & 7.9 & \textbf{10.2} & 7.5 \\
Irish Election Data,\#2 &12& 64 &  \textbf{2.2} & 28.7 &  1.4 & 11.8 &  1.2 & 7.5 &  1.4 & 5.0 &  1.5 & 17.6 &  5.7 & 17.4 &  8.8 & 13.4 & \textbf{20.2} & 12.5 \\
Glasgow City Council,\#7& 13 &32 &  \textbf{1.1} & 9.0 &  0.9 & 3.9 &  0.8 & 2.6 &  1.0 & 1.9 &  1.1 & 9.7 &  1.3 & 9.7 &  1.6 & 9.7 & \textbf{15.8} & 9.7 \\
Glasgow City Council,\#7 &13 &64 &  1.6 & 16.0 &  1.0 & 6.7 &  0.9 & 4.4 &  \textbf{4.6} & 3.2 &  1.6 & 17.0 &  1.9 & 17.0 &  2.6 & 16.8 & \textbf{20.7} & 14.6 \\
ERS Data,\#4 &20 &32 &  \textbf{2.5} & 19.5 &  1.8 & 8.1 &  1.6 & 5.3 &  1.5 & 2.5 &  1.6 & 9.0 &  1.8 & 9.0 &  2.2 & 9.0 & \textbf{2.8} & 9.0 \\
ERS Data,\#4 &20 & 64 &  \textbf{3.7} & 34.0 &  2.3 & 13.9 &  2.0 & 8.8 &  2.0 & 4.1 &  2.1 & 15.8 &  2.5 & 15.8 &  2.8 & 15.8 & \textbf{15.6} & 15.3 \\
\hline
\end{tabular}
\caption{The average time to find an optimal manipulation (avg t) and the average number of manipulators (avg p). Timeout is 1000 sec. \label{table:res}}
\end{table*}
We analysed of the partiality
of voting in real world data sets. We analysed 
the following data sets from PrefLib
\cite{preflib}: Irish Election,  Debian Project, Electoral Reform Society (ERS), Glasgow City Council, F1 and Skiing
and Sushi. In many elections, more than half of the votes contain
less than half of the candidates. Therefore, manipulators
have to deal with partial votes. For each set, we picked several instances and generated 100 elections with $t$ randomly picked votes from the set of votes in the benchmark, where  $t \in \{32,64\}$. On top of this, we vary the length of manipulators' votes.
For each problem instance, we computed the optimal manipulation with a timeout of 1000 sec.
Table \ref{table:res} summarizes our results.

We partition instances into two groups. The first group contains instances with up to 20 candidates. The second group contains
instances with more than 54 candidates. Based on the size of the candidate list, we varied the length of the manipulators' votes differently
in these groups. In the first group the lengths of manipulators' votes are $3,6$ or $9$ and, in the second group,  they are  $15,30$ or $45$. Then we computed the average time and the average number of manipulators in the optimal manipulation over solved instances with partial
or full votes.  It can be seen from the table that there is little correlation between complexity in practice
of finding optimal manipulation with partial and full votes for  Borda$\mbox{}_{\uparrow}$. On the other hand, for the  modified Borda count,
finding an optimal manipulation with full votes is slightly more expensive. The number of manipulators decreases as
the length of the manipulators' votes increases for  Borda$\mbox{}_{\uparrow}$. In contrast, it stays within a 15\% corridor in many benchmarks with the  modified Borda count.

\section{CONCLUSIONS}

In many elections, voters can cast
partial votes. We have studied three of the most
common methods used to modify voting rules
to deal with such partial votes.
These methods modify scoring rules,
elimination rules and rules based on the tournament graph
respectively. We argued that partial voting
may not change the situations where strategic
voting is possible (e.g. with STV). However,
with the Borda count and Copeland's method, partial voting increases
the situations where strategic voting is possible.
As a consequence,  the computational complexity of
computing a manipulation can change.
For example, with the Borda count, the complexity
can decrease or stay the same depending
on how we score partial votes.
We were even able to demonstrate a situation
where the computational complexity of
computing a manipulation increases when
we permit partial voting.

Our results are worst-case and may not reflect
the difficulty of manipulation in practice. A number of
recent theoretical and empirical results
suggest that manipulation can often be computationally
easy on average
(e.g. \cite{csaaai2006} - \cite{wjair11}).
Our NP-hardness results should therefore be seen
as just one of the first steps in understanding the impact
of partial voting on the computational
complexity of computing a manipulation.
There are many other interesting directions to follow.
For example, do results like
these suggest which is the best way to deal with
partial voting? Might we increase our bias for
STV over the Borda count
based on its resistance to manipulation by partial voting.
As a second example, how does partial voting
impact on computational issues surrounding
related problems like possible and necessary winners, control and
bribery?

\bibliographystyle{splncs} 


\end{document}